# What's Fit To Print: The Effect Of Ownership Concentration On Product Variety In Daily Newspaper Markets

August 1, 2001

Lisa George
Department of Economics
Michigan State University

## Abstract

This paper examines the effect of ownership concentration on product position, product variety and readership in markets for daily newspapers. US antitrust policy presumes that mergers reduce the amount and diversity of content available to consumers. However, the effects of consolidation in differentiated product markets cannot be determined solely from theory. Because multi-product firms internalize business stealing, mergers may encourage firms to reposition products, leading to more, not less, variety. Using data on reporter assignments from 1993-1999, results show that differentiation and variety increase with concentration. Moreover, there is evidence that additional variety increases readership, suggesting that concentration benefits consumers.

I am grateful to Joel Waldfogel for many thoughtful comments on earlier drafts. Rachel Croson, Dennis Yao, Julie Wulf, Felix Oberholzer-Gee and participants in the Wharton Applied Economics Seminar also provided useful input. I thank Joshua Katz for generous assistance in obtaining and using data compiled by *Burrelle's Information Services*. All errors are my own. Address correspondence to: Lisa George, Department of Economics, Michigan State University, 101 Marshall Hall, East Lansing, MI 48824, Phone: (517) 355-7583.

# Introduction

Regulation of media markets in the U.S. historically emphasized content and content variety rather than consumer or advertiser prices. This focus accompanies a strong presumption that larger numbers of owners and products in a market lead to better outcomes. Limits on the number of radio stations in a market owned by a single firm, protection of newspaper joint operating agreements, a prohibition against cross-ownership of broadcast and print media products in a market, and active antitrust enforcement against newspaper mergers all attest to this presumption.[1]

Yet, it is far from obvious that more owners give rise to greater variety. Media markets offer differentiated products produced with large fixed costs and advertiser finance. It is well known that such markets can deliver too much, or too little, variety.[2] Duplication can arise if revenue from capturing only a fraction of one consumer type covers fixed costs (Steiner, 1952). At the same time, markets can fail to provide specialized coverage when advertising revenue obtained from targeting a particular group is less than the cost of developing new content (Spence and Owen, 1977). Because multi-product firms internalize business-stealing externalities, mergers can lead owners to eliminate duplicative products and change the content of others. Various production economies, as well as higher revenues, brought about by consolidation can also make new content viable. The effect of ownership concentration on content variety is therefore an empirical question.

---

[1] Restrictions on radio station ownership were relaxed significantly by the Telecommunications Act of 1996.

[2] The tendency for differentiated product markets to produce too much or too little variety is considered in Spence (1976a, 1976b) and Dixit and Stiglitz (1977).

This paper examines the effect of ownership concentration on product position, product variety, and readership in markets for daily newspapers. Newspapers provide a useful setting for studying the effects of concentration on variety for several reasons. First, policy interest in newspapers and newspaper content in particular demonstrate the importance of understanding factors that lead to greater variety. Second, high fixed costs and other aspects of newspaper production limit to a handful the number of products available in any market, raising the consequences of positioning decisions by individual firms.[3] Newspaper owners also appear to have little scope for price discrimination that might otherwise allow for provision of content demanded by small groups.[4] Finally, the 1990's saw a sharp increase in newspaper mergers and acquisitions, driven in part by exogenous policy changes. This consolidation wave aids empirical investigation of the effect of concentration on variety.

This study uses newspaper-level information on the assignment of reporters and editors to approximately 150 different topical reporting beats in 1993 and 1999 to characterize the separation between products and amount of content variety available in 207 newspaper markets. Using a simple distance formula to measure the degree of differentiation in coverage among papers in each market at the beginning of the consolidation wave and six years later, results show that a decrease in the number of owners in a market leads to an increase in separation between products. Moreover, the number of topical reporting beats covered per market also increases with ownership concentration. Finally, there is evidence that the additional coverage brought about by consolidation increases readership. Although policy may be concerned with aspects of diversity beyond the number of topics covered by newspapers, these results suggest that from the standpoint of variety, increased concentration does not harm readers.

---

[3] See Reddaway (1963) and Rosse (1970) for background on the cost structure of the newspaper industry. Advertising accounts for about 80% of newspaper revenue. For estimates in the literature see Compaine (1980, 1982) and Picard (1988). The Newspaper Association of America's *Facts about Newspapers (2000)* reports circulation revenue for daily and Sunday papers at $10.5 billion and advertising revenue at $46.2 billion. See www.naa.org/info/facts00. On average, about 8 newspapers and 6 owners operate per market in the US.

[4] According to data in *Burrelle's Media Directory*, the primary data source for this study, there is little variation in newspaper prices, with more than 75% of daily newspapers selling at $.50 in 1999. Burrelle's data also reveal no relationship between prices and content variety, as measured by the total number of reporting beats covered by the paper. High fixed costs also seem to make the proliferation of tailored content through specialized newspapers very costly, since only the largest markets can support legal, industry, or business dailies.

The paper proceeds as follows. Section 1 reviews the policy and theory that motivate the study. Section 2 examines the relevant literature. Section 3 outlines the data and content measures used in the analyses. Section 4 describes the empirical strategy and Section 5 presents results. Section 6 concludes the paper.

## 1. Motivation

U.S. antitrust policy toward newspapers and regulation of media markets focuses generally on content rather than consumer or advertiser prices. Recent antitrust actions against newspaper mergers and acquisitions in Arkansas, Hawaii, and California are all based on the premise that more papers and more owners in a market lead to greater variety and diversity of content.[5] In Arkansas, the Department of Justice argued that despite their location in different cities, competition between *the Northwest Arkansas Times* and the *Morning News of Northwest Arkansas* led publishers to introduce new content that benefited readers.[6] In Hawaii, the

---

[5] In Arkansas, see in particular U.S. request for oral argument, *U.S.* v. *NAT, L.C.* 139 F.3d 1180 (1998). In Hawaii, see Brief Amicus Curiae of the United States of America in Support of Appellee State of Hawaii and Affirmance, *State of Hawaii* v. *Gannett Pacific Corp.* 203 F.3d 832 (1999) and *State of Hawaii* v. *Gannett Pacific Corp.* 99 F. Supp. 2d 1241 (1999). A brief outline of antitrust concerns in the Hearst case can be found in Department of Justice press release "Hearst Corp. To Sell San Francisco Examiner to ExIn LLC, Resolves Justice Department's Antitrust Concerns," March 30, 2000, available at www.usdoj.gov/atr.

[6] See U.S. request for oral argument, *U.S.* v. *NAT, L.C.* 139 F.3d 1180 (1998). In its appellate brief the DOJ describes the nature of competition between the papers: "The *Times* increased its use of color, and the *Morning News* responded with additional use of color (GX 19 at TC 004827, GA 212). Smith distributed free copies of the *Times* to *Morning News* readers on holidays (GX 99 at NAT08-00101, A 632; GX 44, GA 215). After the *Times* did so in Springdale, the *Morning News* switched to 365-days a year publication (GX 45, GA 216; T. 1251-53, GA 146-48). The *Morning News* followed the *Times* in introducing a travel page (GX 119 at NAT07-00080, GA 282), and the *Times*, in turn, improved its weather page in response to similar improvements by the *Morning News* (GX 231, GA 346). Perhaps most visibly,

Department argued on appeal that early termination of a joint operating agreement allowing the *Honolulu Star-Bulletin* to close would adversely affect news content in the market. Quoting a lower court ruling, the Department argued:

> [N]o monetary amount will be able to compensate for the loss of the *Star-Bulletin's* editorial and reportorial voice, the elimination of a significant forum for the airing of ideas and thoughts, the elimination of an important source of democratic expression, and the removal of a significant facet by which news is disseminated in the community.[7]

In California, antitrust concerns about a transaction that would allow the Hearst Corporation, owner of the *San Francisco Examiner*, to buy the *San Francisco Chronicle* were related similarly to the effect of ownership concentration on newspaper content.[8] Federal Communication Commission regulations prohibiting cross-ownership of newspapers and radio stations in the same market are also based on maintaining diversity.[9]

Despite the concern that newspaper mergers and acquisitions reduce the amount of content available to consumers, several features of newspaper markets make theoretical predictions about the effects ownership concentration unclear. Because differentiated products are somewhat substitutable for one another, they tend to divert or "steal" business from each other. With atomistic ownership and fixed costs, there can be a tendency for duplicative

---

the *Times* expanded its sports coverage in order to "rattle the [Morning] News' cage" (GX 26 at TC 001961, PA 829), and the *Morning News* responded by enhancing its own sports coverage (GX 129, at NAT07-00048, GA 314). . . . In sum, as a *Morning News* official put it, competition in the region was 'a ferocious dog eat dog situation.' (PX 39 at DONR-10708, GA 361)."

[7] Brief Amicus Curiae of the United States of America in Support of Appellee State of Hawaii and Affirmance, *State of Hawaii* v. *Gannett Pacific Corp*. 203 F.3d 832 (1999). The appellate brief quotes the district court ruling, *State of Hawaii* v. *Gannett Pacific Corp*. 99 F. Supp. 2d 1241 (1999).

[8] See Department of Justice press release "Hearst Corp. To Sell San Francisco Examiner to ExIn LLC, Resolves Justice Department's Antitrust Concerns," March 30, 2000, available at www.usdoj.gov/atr.

[9] See 50 FCC 2d at 1074. According to the FCC, rules prohibiting ownership of newspapers and radio stations in the same market are based on "the twin goals of promoting diversity of viewpoints and economic competition." The Commission is currently considering changes to the rule. See FCC-00-191, *1998 Biennial Regulatory Review – Review of the Commission's Broadcast Ownership Rules and Other Rules Adopted Pursuant to Section 202 of the Telecommunications Act of 1996.*

products.[10] Joint production of multiple products, sometimes brought about by mergers, internalizes this business-stealing externality. A newspaper owner who acquires a local competitor's paper similar to his own would likely not continue to operate both papers in their previous forms. Rather, the owner would either differentiate them by altering their content, or close one of the papers altogether. Differentiation might take the form of simply eliminating duplicative content from one paper, replacing duplicative content with new material, or shifting emphasis among reporting topics. In closing one of the papers, the owner might add content to the remaining product to prevent competitor entry into a formerly viable niche. If the acquisition allowed the owner to increase advertising revenue per reader, the owner might also introduce content previously unavailable in the market.[11] Hence even mergers that reduce the number of newspapers would not necessarily reduce content variety. The effect of ownership concentration on variety is an empirical question.

Although theory alone cannot predict the effects of concentration, empirical tests are complicated by the need for exogenous changes in ownership concentration. While there has been no single major reform in newspaper ownership rules in recent years, the 1990's saw considerable increases in ownership concentration across markets driven in part by exogenous factors. Figure 1 shows the number of newspaper acquisitions each year since 1980. In all but one case the number of newspapers changing hands each year since 1993 has exceeded the number sold each year since 1980. Unlike transactions in earlier decades, most transactions in the 1990s have been among ownership groups seeking geographic consolidation.[12] Between the start of the boom in 1993 and 1999, the number of owners per market decreased by about 9% overall and close to 15% in the largest markets.[13]

---

[10] This is one of the possibilities outlined theoretically by Spence (1976a,1976b) and Dixit and Stiglitz (1977). See Berry and Waldfogel (1999) for empirical evidence from another media industry, radio broadcasting.

[11] Although the prediction that multi-product firms seek to separate products is unambiguous, there has been disagreement in the theoretical literature as to whether total product variety ever increases with concentration. In media markets, duplication arises if advertising revenue from capturing only a fraction of one consumer type covers fixed production costs (Steiner, 1952). Beebe (1977) points out that a monopolist would eliminate duplication but would have no incentive to replace it with new content. The monopolist's decision to produce content is considered in Anderson and Coate (2000). In their model of broadcast markets, although a monopolist resists adding content that would lower consumption of existing products, the fact that the monopolist can earn more revenues per program creates an incentive to generate new content. The dominant effect depends on programming costs, the value of consumers to advertisers, the nuisance cost of advertising, and the substitutability of products.

**Figure 1: Newspaper Acquisitions per Year, 1980-1999**

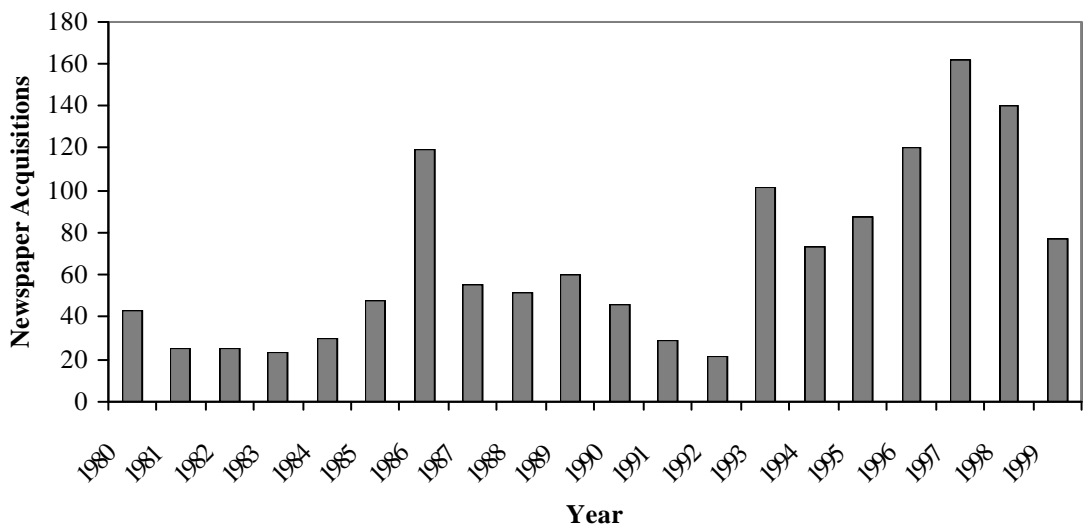

The consolidation wave was fueled by regulatory changes in media ownership rules. In 1993 Congress eliminated restrictions on the FCC's ability to amend rules prohibiting newspaper owners from operating broadcast stations in the same market. In the same year, the FCC granted the first permanent waiver of newspaper cross-ownership rules to Fox Television Stations, Inc., allowing cross-ownership of WNYW-TV and the *New York Post*. The Telecommunications Act of 1996 relaxed limits on the number of jointly-owned radio stations that could be licensed per market, unleashing a wave of consolidation in radio that further altered incentives for media firms to own newspapers *vis a vis* broadcast products across markets.[14]

---

[12] The fraction of independent newspapers (papers publishing in a single city) to all daily newspaper sold declined from 70% in 1985 to fewer than 10% in 1996. Transaction details are taken from Dirks, Van Essen & Associates, "Near-Record Number of Daily Newspapers Sold in 1996," *Newspaper Acquisitions*, Santa Fe, p. 2, 1997 and Dirks, Van Essen & Associates, "Clustering: Growing Regional Groups Retaining Readers as Industry Circulation Slips," *Newspaper Acquisitions,* Santa Fe, p. 2, 1998. Note that the 1986 spike is due to tax changes that induced many owners of independent newspapers to sell at that time.

[13] Concentration measures are calculated from data published in *Burrelle's Media Directory*, the primary data source for this study.

## 2. Literature

Although there is an extensive theoretical literature on the effects of concentration on product position and product variety, there is relatively little empirical evidence on the question.[15] Most closely related to this study is Berry and Waldfogel (2001), which examines the effects of ownership concentration on programming variety in radio broadcasting. They find that consolidation triggered by the Telecommunications Act of 1996 reduced entry but increased the number of radio formats broadcast both absolutely and relative to the number of stations.

In addition to economic work, there is a substantial literature in sociology and communications that examines the relationship between ownership and content in media industries.[16] Much of the work on newspapers focuses on differences between papers published by chains and those published by independent owners, with few clear, robust results. Lacy (1991), one of the more extensive studies in this literature, finds that chain-owned papers contain shorter articles and devote less space to news and editorial beats than independently owned papers, but that they also devote more staff resources for a given allocation of space. Hicks and Featherstone (1978) find evidence that newspapers under common ownership tend to produce less overlapping content. Another branch of this literature examines the effect of ownership structure on political endorsements, again with conflicting results.[17] While most of these studies consider small samples with cross-sectional data, they highlight a long-standing concern with ownership and content reflected in public policy. Because of the close relationship between newspapers and political information, the effects of ownership concentration on content are of interest to a wider audience than might otherwise be the case.

---

[14] For background on FCC cross-ownership restrictions for newspapers see FCC-00-191, *1998 Biennial Regulatory Review – Review of the Commission's Broadcast Ownership Rules and Other Rules Adopted Pursuant to Section 202 of the Telecommunications Act of 1996.* For information on how changes in broadcasting affect newspapers, see *Emergency Petition for Relief of The Newspaper Association of America,* August 23,1999, available at www.naa.org/ppolicy/govt/fccpetition. Economic growth in retail sales beginning in the early 1990's also created incentives for newspaper consolidation. Since retail advertising constitutes about one half of all newspaper advertising revenue, market power in the supply of advertising becomes particularly attractive in active retail markets. For a general discussion of the relationship between competition and advertising see Becker and Murphy (1993).

[15] Hotelling (1929) provides a foundation for the literature on product positioning. Spence (1976a, 1976b) and Dixit and Stiglitz (1977) form the basis of the theoretical literature on product variety.

[16] See Compaine (1995) for a useful review of the communications literature on ownership and content.

This paper also builds on a substantial literature on the industrial organization of media markets generally and the effects of concentration in particular. Early papers by Reddaway (1963) and Rosse (1970) establish the importance of fixed costs in newspaper production. Chaudhri (1998) and Blair and Romano (1993) examine the pricing decisions of a newspaper monopolist earning revenues from circulation and advertising. Both find that features of the newspaper market can lead to lower consumer prices with monopoly ownership. Empirically, Bucklin, Caves, and Lo (1989) find no evidence of a relationship between concentration and circulation or advertising prices, although Reimer (1992) provides evidence from a very small (35 city) sample that concentration may lead to lower prices for advertisers. Dertouzos and Trautman (1990) examine the effect of concentration on firm costs, finding no clear evidence of scale economies from consolidation. A final line of research uses newspaper firms as an avenue for the study of technology adoption and diffusion (Dertouzos and Quinn,1985; and Genesove, 1999).

---

[17] Studies of chain ownership and political bias are numerous. See Akhavan-Majid et al. (1991), Gaziano (1989), Olien et al. (1988) and Busterna and Hansen (1990).

## 3. Data Sources and Content Measures

*3.1 Data Overview*

The basic data set used in this study is a newspaper-level panel identifying the topical beat assignments of reporters and editors at about 1,500 daily newspapers for the years 1993 and 1999. Data from 1995 are also available and used in some specifications. The data also include the owner of each newspaper and total circulation in each year. Newspaper data are supplemented with zip code-level newspaper circulation, population data from the 1990 census, and retail concentration measures from the 1997 economic census. The distribution of reporters across beats is used to construct market-level measures of product differentiation and content variety for 207 designated market areas (DMA's) each year.[18]

Newspaper reporter data come from the 1994, 1996, and 2000 editions of *Burrelle's Media Director* with data reported in 1993, 1995, and 1999.[19] The 2000 edition of the directory maps the job title of about 30,000 reporters and editors (e.g., "Travel Editor") into about 150 topical reporting beats. Data from 1994 and 1996 report job titles only, which are linked individually to beat codes in the 2000 *Directory*. Zip code-level circulation data are taken from the Audit Bureau of Circulations (ABC) *Circulation Data Bank (1999)*. ABC is a membership organization that sets standards for reporting newspaper circulation and audits publisher statements for use by advertisers. The data reflect paid newspaper sales at newsstands, by home delivery, and by mail for 1995 and 1999.[20]

---

[18] Designated Market Areas (DMA's) are delineated by Nielsen Media Research and commonly used by media firms in measuring market participation and compiling consumer demographics for advertisers. DMA's generally correspond to Metropolitan Statistical Areas (MSA's) used in census reporting. However DMA's cover virtually all of the United States, unlike MSA's which exclude rural areas.

[19] *Burrelle's Media Directory* is a product of Burrelle's Information Services, a media monitoring organization that publishes directories and broadcast transcripts, disseminates news clippings, and performs a range of public relations and advertising support functions related to media monitoring.

*3.2 Variety, Differentiation, and Concentration Measures*

An ideal measure of content variety would reflect the number and types of articles published across newspapers over time. Although studies using such output measures are common in the communications literature, the difficulty of assembling data from individual newspapers typically leads to cross-sectional analyses with very small samples in few reporting categories.[21] The basic measure of content variety used in this paper, which allows for more complete characterization of coverage over time using a comprehensive set of daily papers, is the number of different topical beats covered by reporters and editors in a market. Although reporter assignments reflect inputs rather than outputs, the notion that more variety in reporter assignments corresponds to more variety in coverage is highly intuitive. Does a market have a travel editor or not? A music critic? A political analyst? The presence or absence of coverage in a particular topical area is directly related to choices available to readers and hence constitutes a reasonable measure of content variety.[22]

---

[20] To maintain comparability between Burrelle's and ABC data, foreign-language, industry, and national newspapers such as *USA Today*, the *Christian Science Monitor,* and the *Wall Street Journal* are not included in the sample.

[21] Content analysis is widely used to study both the amount of coverage in particular areas as well qualitative features of this coverage. Typical examples are Coulson and Lacy (1998) for newspapers and Hillve, Majanel, and Rosengren (1997) for television.

[22] Beat assignments might also be interpreted as the level of expertise brought to bear on a given news item. The presence or absence of, say, an environmental reporter affects the amount of environmental coverage in a paper. The presence of such a reporter might also improve the quality of coverage of an environmental story that is sufficiently newsworthy to warrant coverage by many papers, whether or not they have an environmental reporter. Although the focus here is on the former effect, both work in the same direction.

Using reporter assignments to measure content variety raises two concerns regarding the link between inputs and outputs. First, Burrelle's assigns reporters to about 150 highly-specific topical beats. With this degree of specificity, reporter assignments might overstate content variety if, for example, a general business editor at a small paper reported on both banking and international trade while the fields were covered by separate reporters at a larger paper. To mitigate potential bias that may be introduced by misconstruing specialization as variety, reporter assignments are aggregated to a set of about 50 consolidated categories.[23] Results in Section 5 are presented for both the entire beat set and for consolidated beats. Second, Burrelle's data include reporter assignments to non-editorial beats such as classifieds, research, and graphics. The link between beat assignments and content is less direct in these non-editorial areas. However, staff assignments in these fields reflect the emphasis of the paper and hence bear directly on product positioning. To capture such differences in emphasis while focusing on the most meaningful beats, results presented for all beat categories include non-editorial beats and results for consolidated beats do not.

Figure 2 shows the average number of reporters per paper assigned to each consolidated beat. Figure 3 shows the percentage of markets in which each beat is covered in 1993 and 1999. General news, sports, national news, entertainment, and opinion are the largest beats covered in all markets by virtually all papers. Topics such as business, food, fashion, and travel are also available in most markets although only at about 20-25% of papers. [24] Computing, gardening, and science reporting are more specialized: coverage is available only in a few percent of papers in about half of the markets. Specialized industry coverage, arts, and multicultural reporting are produced only by a small number of reporters in the very largest markets. No clear trends in coverage are apparent from the graph. Several new beats appear in 1999, such as fine art and multicultural reporting, but the fraction of markets covering more common beats such as banking and consumer affairs declines. The full list of beats is provided in the appendix.

---

[23] Another way to reduce biases that might result from overlapping beat categories is to eliminate columnists and reporters from the underlying beat data and measure content variety and differentiation only on the beat assignments of editorial staff. The effect is similar to consolidation in that editors cover fewer and more general beats than are covered in the combined reporter and editor data. (Markets at the 95th percentile cover 55 beats with editors only in 1999 as opposed to 71 beats and 47 consolidated beats with reporters and editors.) Editorial assignments also appear to be more consistent over time and across papers, likely reducing error or bias that might occur if newspapers with different characteristics use different criteria for categorizing writers or if the degree of inclusion has changed over time. In practice, results using only editorial staff are quite similar to results for consolidated beats and are therefore not presented.

**Figure 2: Average Reporters per Newspaper, Consolidated Beats (1999)**

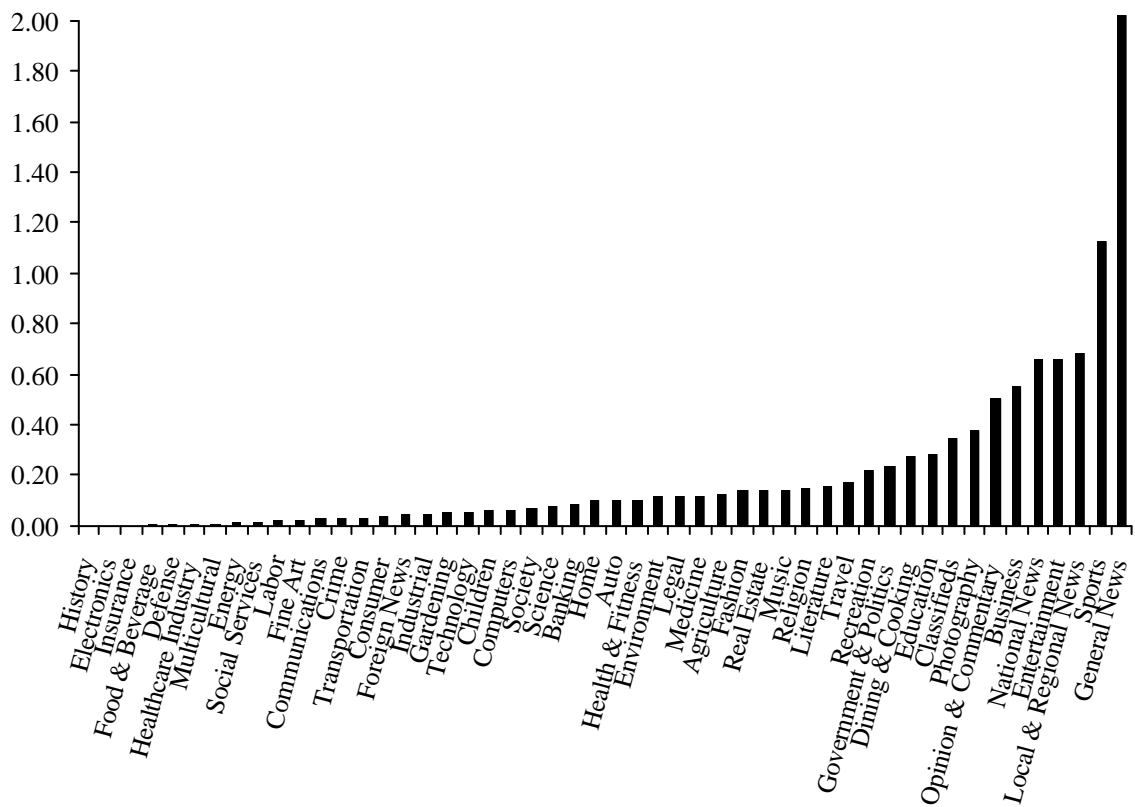

---

[24] Figure 2 shows the total number of reporters assigned to each topic divided by the total number of newspapers in the sample, hence the figure may be interpreted as the average number of reporters per paper or the fraction of papers with a single reporter assigned to the topic. Reporters and editors are often assigned to more than one beat. In these cases beat assignments are calculated proportionally so that a reporter assigned to entertainment and news would be counted as one half of one reporter in each category.

It is also interesting to explore the differences between large and small papers. Figure 4 shows the allocation of reporters to beats at the smallest 20% and largest 20% of daily newspapers. Staff allocations differ significantly, with small papers devoting a much larger fraction of resources to basic topics such as general news, sports, and classifieds. Larger papers assign a greater fraction of reporters to specialty topics such as food, literature, real estate, and politics. These differences suggest that, to the extent mergers and acquisitions produce larger papers, the number of topics covered may increase.

**Figure 3: Percent of Markets Covered, Consolidated Beats (1993, 1999)**

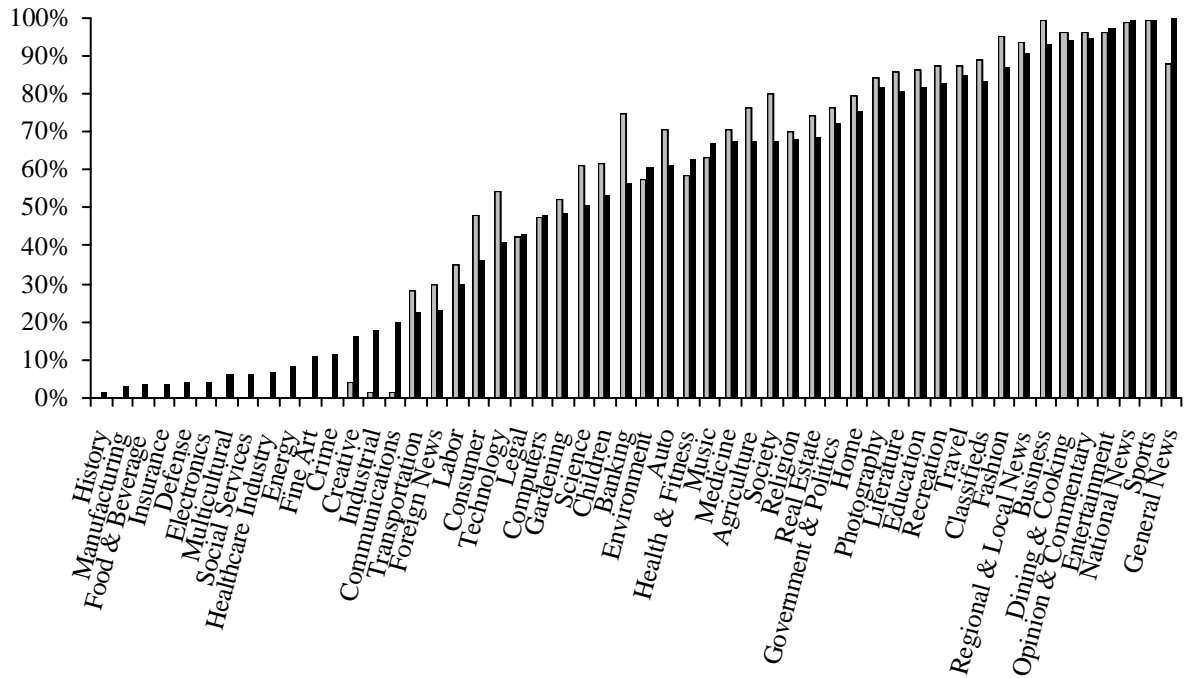

**Figure 4: Top Beat Shares at Large and Small Papers, 1999**

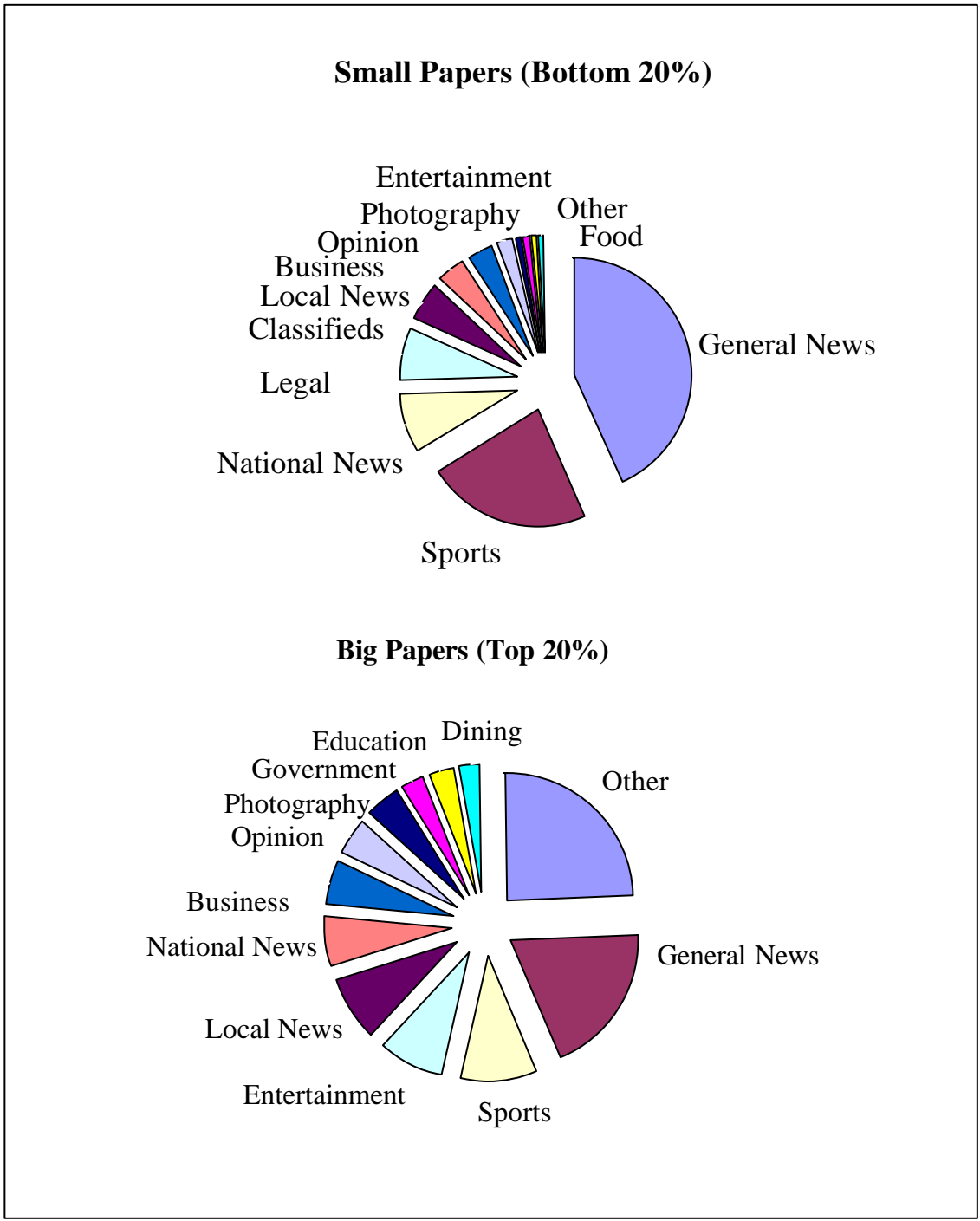

In addition to measuring content variety on an absolute scale, the distribution of reporters is used to compute a measure of the distance between, or differentiation among, products in each market. The change in the measure following consolidation captures whether or not mergers induce firms to spread products apart in product space. For markets with two papers, the measure is simply half the Euclidean distance between them. In markets with more than two papers, it is the average distance to the mean assignment:

$$d = \frac{1}{P}\sum_{i=1}^{P}\sqrt{(s_i - \bar{s})'(s_i - \bar{s})}$$

where $s_i = (s_{i1}, s_{i2}, \ldots, s_{iB})$, $s_{ib}$ is the share of reporters assigned to beat $b$ at paper $i$, $P$ is the total number of papers in the market, and $\bar{s}$ is a vector identifying the average share of reporters assigned to each beat across all papers in a market.[25]

To illustrate the intuition behind the distance measure, consider two markets with two papers assigning reporters to Sports, News, and Entertainment. In the first market, assume that both papers assign one reporter to each beat. In the second market, consider the case where one newspaper assigns one reporter to each beat and the other assigns two reporters to sports and one to news. Differentiation in the first market is zero. In the second market, the average share of reporters assigned to Sports, News, and Entertainment is $\left(\frac{1}{2}, \frac{1}{3}, \frac{1}{6}\right)$. Differentiation in the

---

[25] This type of multi-attribute distance measure is most often found in utility specifications for differentiated products such as in Feenstra and Levinsohn (1995) and Anderson, de Palma, and Thisse (1989).

market is then:

$$d = \frac{1}{2}\left(\underbrace{\sqrt{\left(\frac{1}{3}-\frac{1}{2}\right)^2+\left(\frac{1}{3}-\frac{1}{3}\right)^2+\left(\frac{1}{3}-\frac{1}{6}\right)^2}}_{paper 1} + \underbrace{\sqrt{\left(\frac{2}{3}-\frac{1}{2}\right)^2+\left(\frac{1}{3}-\frac{1}{3}\right)^2+\left(0-\frac{1}{6}\right)^2}}_{paper 2}\right) = \frac{\sqrt{2}}{6}.^{26}$$

The primary independent variable in the paper is ownership concentration, measured as the absolute number of owners in a market and as the number of "owner-equivalents." The number of owner-equivalents is the inverse of the Herfindahl index, defined as the inverse squared sum of ownership market shares, $\frac{1}{\sum_{o=1}^{O} sh_o^2}$, where $sh_o$ is the share of owner $o$ in a market. (Newspaper equivalents are similarly calculated with newspaper rather than ownership shares.) When owners in a market are equal in terms of circulation, the number of owner-equivalents is equal to the number of owners. When circulation is not equal across owners, the number of owner-equivalents is less than the number of owners. In general, owner-equivalents are a better measure of concentration in markets where circulation shares are not equal across firms. For this study, however, using the absolute number of owners provides some advantages. In particular, the number of owner-equivalents depends on newspaper circulation, which is measured with considerably more error than the number of owners or papers in these data.[27] Also, because of its

---

[26] The distance measure could also be calculated based on the number rather than fraction of reporters and editors assigned to each beat. Results are substantively similar to fraction results and are not presented. More direct measures of exclusivity such as the number or fraction of beats in a market covered by a single paper are also possible. Again, results are consistent with those shown and are not presented. Note that 18 of the markets in the sample include only one paper. Measured distance is zero in these markets in both 1993 and 1999. Because markets with a single daily paper have seen no ownership changes, the zeros do not affect estimated coefficients. Their effect on standard errors is also very small.

[27] Specifically, Burrelle's does not disaggregate circulation of each newspaper within and outside the home market and therefore tends to overstate market circulation. Data available from the Audit Bureau of Circulations allows calculation of in-market and out-of-market circulation for some papers. The ratio of out-of-market circulation to total circulation ranges from zero to more than 40%. For example, out-of-market circulation is about

dependence on circulation, the number of owner-equivalents in a market changes even when the number of owners does not, introducing further error into the study of ownership changes.

**Table 1: Summary Statistics**

|  | N | Mean | SD | 5% | 25% | 50% | 75% | 95% |
|---|---|---|---|---|---|---|---|---|
| *Newspaper Statistics (1993)* | | | | | | | | |
| Beats Covered | 1630 | 15.40 | 9.21 | 3 | 7 | 11 | 19 | 35 |
| Consolidated Beats Covered | 1630 | 13.82 | 8.45 | 3 | 7 | 12 | 19 | 31 |
| *Newspaper Statistics (1999)* | | | | | | | | |
| Beats Covered | 1546 | 14.37 | 11.01 | 4 | 8 | 13 | 21 | 35 |
| Consolidated Beats Covered | 1546 | 12.62 | 8.65 | 3 | 6 | 10 | 17 | 31 |
| *Market Statistics (1993)* | | | | | | | | |
| Beats Covered | 207 | 32.53 | 10.43 | 14 | 25 | 34 | 42 | 45 |
| Consolidated Beats Covered | 207 | 28.99 | 8.79 | 13 | 23 | 30 | 37 | 39 |
| Total Staff | 207 | 82.60 | 78.87 | 14 | 31 | 60 | 105 | 238 |
| Owners | 207 | 6.14 | 4.90 | 1 | 3 | 4 | 9 | 16 |
| Owner-equivalents | 207 | 2.77 | 1.64 | 1.00 | 1.64 | 2.32 | 3.43 | 6.39 |
| Papers | 207 | 7.87 | 7.15 | 1 | 3 | 5 | 11 | 23 |
| Paper Equivalents | 207 | 3.24 | 2.06 | 1.00 | 1.81 | 2.75 | 4.08 | 7.10 |
| Distance — All Beats | 207 | 18.04 | 7.11 | 0.00 | 15.99 | 18.85 | 22.01 | 26.56 |
| Distance — Consolidated Beats | 207 | 1.58 | 0.78 | 0.00 | 27.72 | 42.31 | 57.12 | 78.51 |
| Per Capita Newspaper Sales (1995) | 196 | 0.18 | 0.05 | 0.09 | 0.15 | 0.18 | 0.21 | 0.26 |
| *Market Statistics (1999)* | | | | | | | | |
| Beats Covered | 207 | 35.47 | 17.93 | 13 | 23 | 33 | 41 | 71 |
| Consolidated Beats Covered | 207 | 28.82 | 10.52 | 11 | 21 | 30 | 36 | 47 |
| Total Staff | 207 | 98.54 | 134.64 | 12 | 27 | 52 | 110 | 308 |
| Owners | 207 | 5.57 | 4.09 | 1 | 3 | 4 | 8 | 14 |
| Owner-equivalents | 207 | 2.60 | 1.45 | 1 | 1.57 | 2.24 | 3.19 | 5.62 |
| Papers | 207 | 7.47 | 6.41 | 1 | 3 | 5 | 10 | 22 |
| Paper Equivalents | 207 | 3.06 | 1.92 | 1 | 1.77 | 2.52 | 3.90 | 6.62 |
| Distance — All Beats | 207 | 20.17 | 7.66 | 0.00 | 15.99 | 18.85 | 22.01 | 26.56 |
| Distance — Consolidated Beats | 207 | 1.79 | 0.87 | 0.00 | 31.88 | 45.39 | 61.33 | 88.38 |
| Per Capita Newspaper Sales | 196 | 0.17 | 0.05 | 0.09 | 0.14 | 0.17 | 0.20 | 0.25 |

3% for the *Philadelphia Inquirer*, about 9% for the *Washington Post*, 11% for the *San Francisco Chronicle* and 19% for the *Des Moines Register*.

Table 1 summarizes product-level beat data as well as market-level measures of variety, product separation, and ownership concentration in 1993 and 1999. The number of papers drops from 1,630 to 1,546 in the same markets over the time period of the study. The average daily paper covers about 15 beats. The smallest papers cover about five reporting beats and the largest about 35. The means for consolidated beats are somewhat lower than for the disaggregated beats. At the market level, the number of beats covered ranges from about 14 at the 5$^{th}$ percentile to more than 70 at the 95$^{th}$ percentile. With consolidated beats, total content variety ranges from about 13 at the 5$^{th}$ percentile to about 45 at the 95$^{th}$ percentile.$^{28}$ The average number of owners per market each year is about 6, with the number of owner-equivalents about 3. The average number of papers is approximately 8 in both years, with the number of paper equivalents also about 3. The mean distance among papers in a market is 20 for total beats and 2 for consolidated beats.

## 4. Empirical Strategy

The goal of the empirical analyses is to measure the effect of ownership concentration on the separation among papers and the number of topical beats covered in a newspaper market. An empirical approach commonly employed in the communications literature is to regress each market attribute on the number of owners and observable market characteristics such as population, income, and education.$^{29}$ Although cross-sectional analyses can in some cases be informative, this approach is vulnerable to the concern that unobserved differences across markets affect both product differentiation and variety, which could bias results.

---

$^{28}$ Recall that reporters and editors are often assigned to more than one beat. This allows the number of beats covered in a market to exceed the number of reporters.

$^{29}$ Cross-sectional methods are common. See, for example, Hicks and Featherstone (1978) and Akhavan-Majid et al. (1991).

The primary method pursued in this study, which avoids these problems, is estimation with longitudinal data and market fixed effects. With observations at two points in time, a dummy variable can be added to the regressions for each market. The effect of concentration on variety is then identified from the relationship between *changes* in concentration and *changes* in variety. If unobserved preferences for newspaper content across markets remain constant over time, this approach produces unbiased estimates of the effect of changes in the number of owners or owner-equivalents on changes in differentiation or variety. All of the relationships explored in Section 5 are estimated using market fixed effects.

Although the overall burst of acquisition activity that began in 1993 was driven by exogenous regulatory and market factors, ownership changes across markets may themselves be endogenously related to preferences for newspaper content.[30] To address this concern, it is useful to identify an instrument for changes in the number of owners. Such an instrument would need to be correlated with changes in ownership and uncorrelated with changes in preferences. One candidate for an instrument is population, since small markets with very few papers simply have less scope for consolidation than larger markets. Another candidate for an instrument is the amount of retail competition in the market. Local retail advertising constitutes about 45% of all newspaper advertising. If demand for newspaper advertising increases with retail competition, market power in the supply of advertising should be more valuable in more competitive retail markets.[31] Both population levels and the amount of retail competition should be uncorrelated with changes in preferences for newspaper coverage.[32]

---

[30] If acquired newspapers differ systematically from newspapers that do not change hands, selection effects may bias both fixed effects and instrumental variables estimates. For example, if acquired papers are less financially sound than others, an increase in the number of topics covered after acquisition may reflect changes in management rather than concentration. Newspaper-level financial data that would allow general tests for selection effects are not available. However, this specific concern can be studied by regressing the absolute and percentage change in circulation and variety over the period 1993-1995 on a dummy variable equal to one for the 98 newspapers that changed hands in 1996 and zero otherwise. Insignificant coefficient estimates on the dummy variable in regressions with and without a market fixed effect indicate that papers changing owners in 1996 experienced no loss in circulation or variety between 1993 and 1995 relative to papers with a single owner over this period.

Market population is generated by aggregating zip code-level population data from the 1990 Census to the DMA level. Measures of retail competitiveness can be constructed from data in the 1997 Census of Business, which reports total sales by retail category by county. The inverse of the Herfindahl index across retail categories produces an estimate of the number of "retail-equivalents" in a market. This measure of retail concentration should be related to the level of competition.[33]

Table 2 summarizes ownership, differentiation and variety measures in 1993 and 1999 by population quintile and by quintiles of retail-equivalents. The data suggest that the number of owners and owner-equivalents has dropped across all market segments, but the drop has been more pronounced in larger markets and markets with a less concentrated retail sector. The relationship appears to be more consistent across population quintiles than retail quintiles. The change in content variety is also higher in larger markets and markets with less retail concentration. As expected, the total amount of content variety is also much greater in big markets. Markets in the top population quintile publish about twice as many beats as smaller markets in 1993 and three times as many beats in 1999. Markets in the top retail quintile publish in about 50% more beats than smaller markets in 1993 and about twice as many beats in 1999.

---

[31] Newspaper advertising shares are reported in the Newspaper Association of America's, *Facts about Newspapers (2000),* available at www.naa.org/info/facts00/09. See Becker and Murphy (1993) for a general discussion of the relationship between competition and advertising.

[32] *Changes* in population, however, might be correlated with changes in preferences. Robustness checks on results consider the effects of population changes using population estimates from Claritas, a marketing firm that compiles product, industry and demographic data. See notes in Section 5.

[33] Specifically, the number of retail-equivalents in a market is defined as the inverse of the squared sales share in each retail category, $Req = 1 \Big/ \sum_{i=1}^{N} ssh_i^2$, where $ssh_i$ is the retail sales share in category $i$ and $N$ is the total number of retail categories. A better measure would estimate concentration at the establishment level, however the Census Bureau does not release data necessary for such a calculation. The cross-category measures reflect sales in 44 five-digit North American Industry Classification System (NAICS) codes in the retail categories 442-446 and 448-453. Summary statistics for retail data are included in the appendix.

**Table 2: Market Characteristics by Population and Retail Quintiles, 1993-1999**

| Quintile | DMA's | Pop (M) | Retail Eq's | Owners | Owner Eq's | Papers | Paper Eq's | All Beats | Combined Beats | Distance All Beats | Distance Combined Beats |
|---|---|---|---|---|---|---|---|---|---|---|---|
| *1993 - Population Quintiles* | | | | | | | | | | | |
| 1 | 42 | 0.16 | 3.78 | 1.93 | 1.56 | 2.26 | 1.77 | 19.48 | 17.95 | 11.52 | 18.50 |
| 2 | 41 | 0.37 | 4.62 | 3.93 | 2.42 | 4.56 | 2.81 | 29.20 | 26.54 | 17.14 | 34.45 |
| 3 | 42 | 0.64 | 5.94 | 4.98 | 3.00 | 6.10 | 3.45 | 34.12 | 30.43 | 19.29 | 42.08 |
| 4 | 41 | 1.14 | 6.65 | 7.80 | 3.33 | 9.93 | 4.09 | 37.73 | 33.44 | 20.98 | 52.47 |
| 5 | 41 | 3.73 | 7.65 | 12.17 | 3.57 | 16.71 | 4.10 | 42.39 | 36.83 | 21.41 | 62.71 |
| *1999 - Population Quintiles* | | | | | | | | | | | |
| 1 | 42 | 0.16 | 3.78 | 1.90 | 1.52 | 2.19 | 1.70 | 18.74 | 17.02 | 13.39 | 21.17 |
| 2 | 41 | 0.37 | 4.62 | 3.71 | 2.30 | 4.46 | 2.69 | 26.37 | 23.51 | 18.73 | 36.21 |
| 3 | 42 | 0.64 | 5.94 | 4.79 | 2.90 | 5.81 | 3.22 | 32.52 | 28.76 | 20.81 | 44.15 |
| 4 | 41 | 1.14 | 6.65 | 7.24 | 3.07 | 9.68 | 3.77 | 38.12 | 32.71 | 22.91 | 56.48 |
| 5 | 41 | 3.73 | 7.65 | 10.34 | 3.22 | 15.37 | 3.95 | 62.07 | 42.37 | 25.17 | 73.86 |
| *% Change 1993-1999* | | | | | | | | | | | |
| 1 | 42 | - | - | -1.2% | -2.3% | -3.2% | -3.8% | -3.8% | -5.2% | 16.2% | 14.5% |
| 2 | 41 | - | - | -5.6% | -4.9% | -2.1% | -4.5% | -9.7% | -11.4% | 9.3% | 5.1% |
| 3 | 42 | - | - | -3.8% | -3.5% | -4.7% | -6.6% | -4.7% | -5.5% | 7.9% | 4.9% |
| 4 | 41 | - | - | -7.2% | -7.9% | -2.5% | -7.9% | 1.0% | -2.2% | 9.2% | 7.6% |
| 5 | 41 | - | - | -15.0% | -9.8% | -8.0% | -3.7% | 46.4% | 15.0% | 17.6% | 17.8% |
| *1993 - Retail Quintiles* | | | | | | | | | | | |
| 1 | 42 | 0.26 | 2.85 | 3.00 | 1.93 | 3.67 | 2.29 | 23.93 | 21.79 | 14.57 | 27.99 |
| 2 | 41 | 0.52 | 4.72 | 4.02 | 2.37 | 4.80 | 2.78 | 29.07 | 26.24 | 16.55 | 32.33 |
| 3 | 42 | 0.77 | 5.90 | 5.36 | 2.75 | 6.83 | 3.20 | 32.67 | 29.33 | 17.91 | 39.73 |
| 4 | 41 | 1.31 | 6.80 | 7.56 | 3.27 | 9.54 | 3.73 | 36.73 | 32.41 | 20.56 | 52.45 |
| 5 | 41 | 3.17 | 8.40 | 10.83 | 3.57 | 14.66 | 4.21 | 40.44 | 35.34 | 20.72 | 57.53 |
| *1999 - Retail Quintiles* | | | | | | | | | | | |
| 1 | 42 | 0.26 | 2.85 | 2.86 | 1.89 | 3.45 | 2.12 | 21.69 | 19.40 | 15.28 | 28.61 |
| 2 | 41 | 0.52 | 4.72 | 3.83 | 2.30 | 4.78 | 2.72 | 28.44 | 25.17 | 19.58 | 37.78 |
| 3 | 42 | 0.77 | 5.90 | 5.19 | 2.66 | 6.60 | 3.03 | 33.00 | 28.07 | 19.82 | 43.04 |
| 4 | 41 | 1.31 | 6.80 | 7.07 | 3.04 | 9.32 | 3.52 | 40.59 | 33.71 | 22.35 | 55.54 |
| 5 | 41 | 3.17 | 8.40 | 9.00 | 3.12 | 13.32 | 3.92 | 54.02 | 37.98 | 23.96 | 66.75 |
| *% Change 1993-1999* | | | | | | | | | | | |
| 1 | 42 | - | - | -4.8% | -2.2% | -5.8% | -7.4% | -9.4% | -10.9% | 4.9% | 2.2% |
| 2 | 41 | - | - | -4.8% | -3.0% | -0.5% | -2.0% | -2.2% | -4.1% | 18.3% | 16.9% |
| 3 | 42 | - | - | -3.1% | -3.2% | -3.5% | -5.4% | 1.0% | -4.3% | 10.6% | 8.3% |
| 4 | 41 | - | - | -6.5% | -6.9% | -2.3% | -5.6% | 10.5% | 4.0% | 8.7% | 5.9% |
| 5 | 41 | - | - | -16.9% | -12.5% | -9.2% | -6.9% | 33.6% | 7.5% | 15.7% | 16.0% |

Notes: Owner, paper and retail-equivalents calculated as the inverse Herfindahl index. See text for details.

Table 3 presents the results of a regression of the change in the number of owners and the change in the number of owner-equivalents on market population and on the number of retail-equivalents. The coefficients on population and retail-equivalents are negative, confirming evidence in Table 2 that newspaper ownership concentration has increased the most in large markets and markets with more retail competition. With both instruments, the amount of variation explained by the owner model is much higher than the model with owner-equivalents.[34] The population specification also explains variation in ownership changes better than the number of retail-equivalents. Given these results, regressions in Section 5 present fixed effects estimates for both the number of owners and number of owner-equivalents and instrumental variables results only for the number of owners.

**Table 3: Instruments for Changes in Ownership Concentration**

|  | Δ Owners (1993-99) | Δ Owners Equivalents (1993-99) | Δ Owners (1993-99) | Δ Owners Equivalents (1993-99) |
| --- | --- | --- | --- | --- |
|  | (1) | (2) | (3) | (4) |
| DMA Population (M) | -0.676 | -0.110 |  |  |
|  | (7.82)** | (2.11)* |  |  |
| DMA Population Squared | 0.019 | 0.005 |  |  |
|  | (3.19)** | (1.36) |  |  |
| Retail-equivalents |  |  | -0.262 | -0.066 |
|  |  |  | (5.72)** | (2.86)** |
| Constant | 0.151 | -0.067 | 0.940 | 0.204 |
|  | (1.39) | (1.03) | (3.39)** | (1.46) |
| Observations | 207 | 207 | 207 | 207 |
| R-squared | 0.39 | 0.03 | 0.14 | 0.04 |

Notes: t-statistics in parentheses (*p<.05, **p<.01). Changes in owners, owner-equivalents 1993-1999 on population and retail-equivalents (levels). Owner-equivalents and retail-equivalents calculated as the inverse Herfindahl index.

---

[34] As noted in Section 3.2, a large amount of measurement error in Burelle's circulation estimates is likely responsible for the limited explanatory power in regressions using owner-equivalents. Note that it is also possible to instrument for the ownership Herfindahl directly, rather than equivalents, its inverse. It turns out that the explanatory power with the Herfindahl is less than for the number of owner-equivalents and retail-equivalents and is not used.

# 5. Results

This section proceeds in two parts. The first examines the relationship between ownership concentration, product positioning, and product variety. The second considers readership.

*5.1 Ownership Concentration, Product Positioning, and Product Variety*

Because multi-product firms are better able to internalize business-stealing, they want their papers to look different from each other and appeal to different readers. One way that newspaper owners might differentiate products is to shift the allocation of reporters assigned to particular beats. Recall from Section 3 that differences in emphasis among papers in a market can be measured as the average Euclidian distance among papers. Table 4 shows the results of fixed-effects and instrumental variables regressions of changes in the distance measure on changes in the number of owners and owner-equivalents and on changes in the number of owners and papers. The first two columns show the effect of an increase in owners and owner-equivalents, the third column shows the effect of changes in owners and papers, and the fourth and fifth columns show the effect of changes in the number of owners using population and retail-equivalents as instruments for changes in owners. Results are repeated for the complete set of reporting beats and for consolidated beats. The relationship is always negative and virtually always significant, and the coefficient estimates are consistent across columns. Overall, a decrease of one owner or owner-equivalent in the market leads to an increase in distance among papers by about 2-4%. The number of newspapers has no effect. Increased ownership concentration, regardless of the number of products, appears to increase the distance between products.

**Table 4: Does Ownership Concentration Increase Separation Among Papers?**

| | Distance Among Papers, 1993-1999 | | | | | | | | | |
|---|---|---|---|---|---|---|---|---|---|---|
| | *All Beats* | | | | | *Consolidated Beats* | | | | |
| | FE | FE | FE | IV Pop | IV Retail | FE | FE | FE | IV Pop | IV Retail |
| | (1) | (2) | (3) | (4) | (5) | (6) | (7) | (8) | (9) | (10) |
| Owners | -0.60 | | -0.69 | -0.79 | -1.02 | -1.73 | | -2.07 | -3.36 | -3.89 |
| | (2.95)** | | (2.55)* | (2.42)* | (1.82) | (2.87)** | | (2.56)* | (3.21)** | (2.17)* |
| Owner-equivalents | | -0.66 | | | | | -3.06 | | | |
| | | (1.46) | | | | | (2.30)* | | | |
| Papers | | | 0.13 | | | | | 0.48 | | |
| | | | (0.52) | | | | | (0.63) | | |
| Constant | 22.63 | 20.89 | 22.16 | 1.68 | 1.55 | 54.23 | 52.30 | 52.53 | 2.42 | 2.13 |
| | (18.7)** | (16.9)** | (14.5)** | (4.9)** | (3.6)** | (15.1)** | (14.5)** | (11.7)** | (2.2)* | (1.5) |
| DMA's | 207 | 207 | 207 | 207 | 207 | 207 | 207 | 207 | 207 | 207 |

Notes: t-statistics in parentheses (*p<.05, **p<.01). Regressions show distance among papers on owners, owner-equivalents, and papers. Constants in FE regressions represent the average value of the DMA fixed effect. IV regressions predict changes in owners with population and retail-equivalents (levels). Constants in IV regressions show average change in the independent variable.

Product positioning is interesting because predictions emerge directly from theory on firm behavior in differentiated product markets. However the effect of concentration on the total amount of variety available in a market offers a more direct link to consumer welfare. As a first step in examining the effect of ownership concentration on total product variety, it is useful to ask how an increase in concentration affects the number of papers in a market. This is done by regressing changes in concentration measures on changes in the number of papers. Table 5 presents results. The first column shows the relationship in terms of changes in the number of owners and papers and the second column shows the effect in terms of owner and paper equivalents. The third and fourth columns show the effect of ownership changes using population and the number of retail-equivalents as instruments for the change in owners. In each case, the relationship between owners and papers is positive and significant, suggesting that increased ownership concentration reduces the number of products available in a market.

**Table 5: Does Ownership Concentration Reduce the Number of Newspapers?**

| | Number of Daily Newspapers, 1993-1999 | | | |
|---|---|---|---|---|
| | Δ Papers | Δ Paper Equivalents | Δ Papers | Δ Papers |
| | FE | FE | IV - Pop | IV- Retail |
| | (1) | (2) | (3) | (3) |
| Δ Owners | 0.701 | | 1.080 | 0.731 |
| | (11.73)** | | (10.32)** | (4.53)** |
| Δ Owner-equivalents | | 0.835 | | |
| | | (16.62)** | | |
| Constant | -0.013 | -0.034 | 0.199 | 0.004 |
| | (0.15) | (0.99) | (1.84) | (0.03) |
| Observations | 207 | 207 | 207 | 207 |
| R-squared | 0.40 | 0.57 | 0.28 | 0.40 |

Notes: t-statistics in parentheses (*p<.05, **p<.01). Changes in paper and paper-equivalents on changes in owners and owner-equivalents, 1993-1999.

A decline in the number of products is a first-order effect that would be expected to decrease variety. However, as discussed above, the full impact of concentration on variety depends on how owners alter products that remain in the market. Turning to the primary question, Table 6 shows the effect of changes in the number of owners and owner-equivalents on changes in the total number of reporting beats covered in a market. The first three columns show the effect of changes in owners, owner-equivalents, and the number of papers using market fixed effects and the fourth and fifth columns show the effect of changes in owners on changes in variety using population and retail concentration as instruments for changes in owners. The final five columns repeat the analyses using consolidated beats. The coefficients on ownership measures are negative and significant in all but one case. On average, the loss of one owner in a market increases the number of reporting beats covered by about 4 beats on a base of about 60 total beats (an increase of about 6%). The loss of one owner leads to an increase of 1 consolidated beat on a base of 30 consolidated beats (an increase of about 3.3%). Results using owner-equivalents show a slightly smaller effect on the number of beats covered. Instrumental variables estimates are somewhat larger.[35] Ownership concentration appears to increase the number of beats covered in a market.

It is worth noting that the results in Table 6 are driven by changes in the number of *owners*. The effect of changes in the number of *papers* on changes in the number of beats covered is negative for the total beat set, although the effect is smaller than the effect of owners. The effect of papers is negligible and insignificant for consolidated beats. No strong conclusions can be drawn from this result. However, it does suggest that mergers produce more total content when one paper closes than when both continue to operate. Since operating multiple papers provides the opportunity to differentiate products through reporting emphasis as well as total content, the result is not unreasonable.

---

[35] Checks on these results show that the magnitude and significance of the effect of changes in ownership concentration on changes in the number of beats covered in a market are robust to inclusion of market demographics (fraction educated, black, young, old) and estimated changes in population between 1990 and 1999.

**Table 6: Does Ownership Concentration Increase Content Variety?**

| | Number of Reporting Beats Covered, 1993-1999 | | | | | | | | | |
|---|---|---|---|---|---|---|---|---|---|---|
| | *All Beats* | | | | | *Consolidated Beats* | | | | |
| | FE (1) | FE (2) | FE (3) | IV Pop (4) | IV Retail (5) | FE (6) | FE (7) | FE (8) | IV Pop (9) | IV Retail (10) |
| Owners | -4.32 (8.94)** | | -3.35 (5.23)** | -10.91 (9.69)** | -9.70 (5.54)** | -0.96 (3.75)** | | -0.97 (2.85)** | -3.26 (6.55)** | -3.27 (3.90)** |
| Owner-equivalents | | -2.607 (2.11)* | | | | | -0.475 (0.82) | | | |
| Papers | | | -1.386 (2.29)* | | | | | 0.025 (0.08) | | |
| Constant | 59.34 (20.7)** | 40.99 (12.2)** | 64.26 (18.0)** | -3.17 (2.7)** | -2.49 (1.8) | 34.53 (22.8)** | 30.17 (19.2)** | 34.44 (18.0)** | -2.00 (3.8)** | -2.01 (3.1)** |
| DMA's | 207 | 207 | 207 | 207 | 207 | 207 | 207 | 207 | 207 | 207 |

Notes: t-statistics in parentheses (*p<.05, **p<.01). Regressions show beats covered on owners, owner-equivalents, and papers. Constants in FE regressions represent the average value of the DMA fixed effect. IV regressions predict changes in owners with population and retail-equivalents (levels). Constants in IV regressions show average change in the independent variable.

This increase in total content variety is surprising particularly in light of the result that an increase in concentration reduces the number of papers in a market, which, by itself, works against more variety. The result suggests that multi-paper firms produce variety more efficiently, covering the same number of beats with fewer papers or more beats with the same number of papers. This "efficiency effect" can be studied directly by examining the effect of changes in ownership concentration on changes in the total number of beats covered in a market divided by the number of papers. Results are shown in Table 7. The coefficients are all negative and all but one significant. A decrease of one owner appears to increase the average number of topics per paper by about 0.3 beats or roughly 3%. The estimates using owner-equivalents are somewhat larger, with a reduction of one owner-equivalent leading to an increase in variety per paper by about 2 beats per paper or about 10%. Concentration appears to lead to more efficient production of variety.

Taken together, these results suggest that although increased ownership concentration reduces the number of papers in a market, the increase in concentration leads to greater separation among papers. Moreover, concentration appears to increase total content variety, providing stronger evidence that newspaper consolidation can benefit readers.

*5.2 Ownership Concentration and Readership*

This section asks whether the additional variety generated from consolidation increases readership. Figure 5 shows the relationship between per capita newspaper sales and total content variety for 1995 and 1999 using zip code-level circulation data from ABC aggregated to the market level.[36] The relationship appears positive, suggesting that additional content can attract

---

[36] Readership data are available for 1995 and 1999 only and are incomplete in some markets. Incomplete markets are excluded from the sample.

**Table 7: Does Ownership Concentration Increase Average Variety per Paper?**

| | Number of Reporting Beats Covered per Paper, 1993-1999 | | | | | | | |
| | *All Beats* | | | | *Consolidated Beats* | | | |
| | FE | FE | IV Pop | IV Retail | FE | FE | IV Pop | IV Retail |
| | (1) | (2) | (3) | (4) | (5) | (6) | (7) | (8) |
|---|---|---|---|---|---|---|---|---|
| Owners | -0.324 | | -0.799 | -1.288 | -0.150 | | -0.404 | -0.857 |
| | (2.54)* | | (3.52)** | (3.11)** | (1.35) | | (2.10)* | (2.49)* |
| Owner-equivalents | | -2.177 | | | | -1.518 | | |
| | | (4.17)** | | | | (4.18)** | | |
| Constant | 8.866 | 19.114 | -0.277 | -0.551 | 7.033 | 15.529 | -0.331 | -0.585 |
| | (11.76)** | (13.54)** | (1.18) | (1.75) | (10.75)** | (15.82)** | (1.66) | (2.23)* |
| DMA's | 207 | 207 | 207 | 207 | 207 | 207 | 207 | 207 |

Notes: t-statistics in parentheses (*p<.05, **p<.01). Regressions show average beats per paper on owners, owner-equivalents, and papers. Constants in FE regressions represent the average value of the DMA fixed effect. IV regressions predict changes in owners with population and retail-equivalents (levels). Constants in IV regressions show average change in the independent variable.

readers to a market. Table 8 shows the relationship between changes in the number of beats covered in a market and changes in per capita newspaper sales. As expected, the effect is positive for both the total beat set and for consolidated beats, although coverage of 10 additional beats in a market appears to increase newspaper sales per capita by only 0.003, a small effect. The regressions in Table 9 estimate the effect of changes in the number of owners and owner-equivalents over time on readership directly. As above, the relationship is estimated with market fixed effects and instrumental variables. The effect of an increase in owners is always negative but significant in only two cases. The coefficient is small, with the loss of one owner raising total per capita circulation by 0.002-0.009 or about 1-5%. Although these results provide only limited evidence that increases in ownership concentration increase readership, there appears to be no evidence that ownership concentration *reduces* readership. Taken together, the results in this section suggest that the new content which emerges from consolidation does not reduce demand for newspapers.

**Figure 5: Per Capita Newspaper Sales and Number of Beats Covered; 1995, 1999**

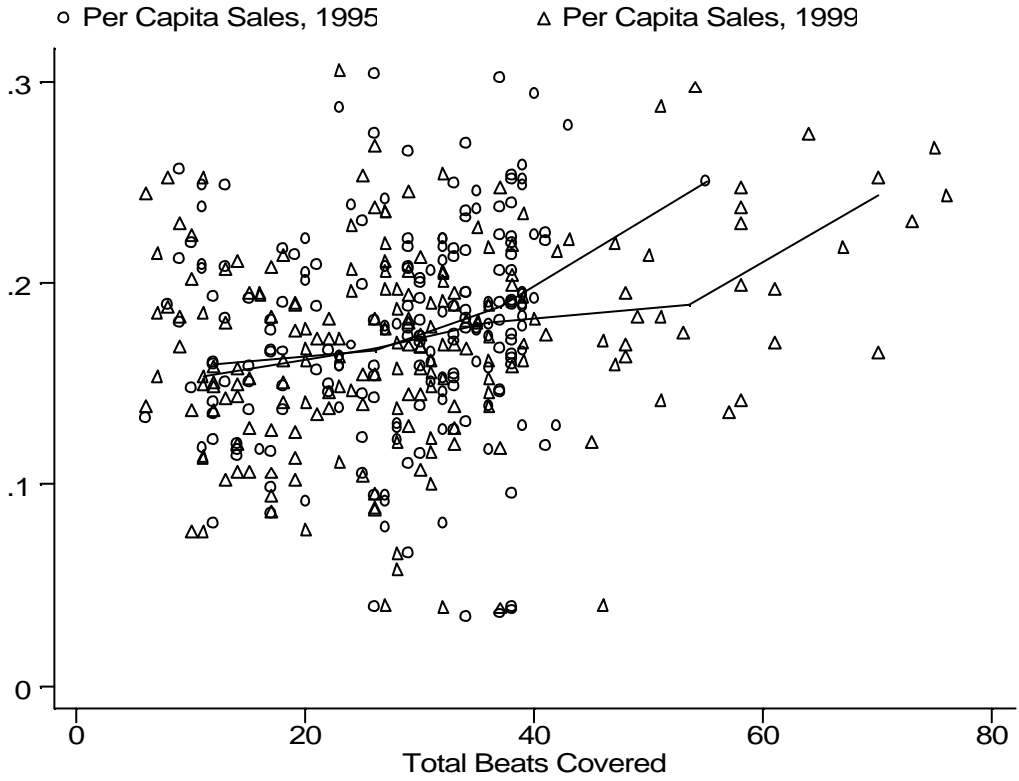

**Table 8: Does Content Variety Increase Readership?**

|  | Per Capita Newspaper Sales, 1995-1999 | |
|---|---|---|
|  | *All Beats* | *Consolidated Beats* |
|  | FE | FE |
|  | (1) | (2) |
| Δ Total Beats Covered | 0.0003 | 0.0003 |
|  | (1.98)* | (2.11)* |
| Constant | -0.0041 | -0.0037 |
|  | (3.12)** | (2.87)** |
| Observations | 196 | 196 |

Notes: t-statistics in parentheses (*p<.05, **p<.01). Readership data from ABC. Excludes the following 11 markets with incomplete data: Birmingham-Anniston, AL; Dallas-Fort Worth, TX; Jonesboro, AR; Joplin-Pittsburg, MO-KS; Knoxville, TN; Los Angeles, CA; Nashville, TN; New York, NY; Oklahoma City, OK; Tulsa, OK; and Juneau, AK.

**Table 9: Does Ownership Concentration Increase Readership?**

|  | Per Capita Newspaper Sales, 1995-1999 | | | | | |
|---|---|---|---|---|---|---|
|  | FE | FE | FE | FE | IV Pop | IV Retail |
|  | (1) | (2) | (3) | (4) | (5) | (6) |
| Owners | -0.0024 |  | -0.0023 |  | -0.0090 | -0.0045 |
|  | (1.98)* |  | (1.44) |  | (3.72)** | (1.05) |
| Owner-equivalents |  | -0.0033 |  | -0.0038 |  |  |
|  |  | (1.12) |  | (1.08) |  |  |
| Papers |  |  | -0.0002 |  |  |  |
|  |  |  | (0.14) |  |  |  |
| Paper Equivalents |  |  |  | 0.0011 |  |  |
|  |  |  |  | (0.26) |  |  |
| Constant | 0.1879 | 0.1829 | 0.1890 | 0.1809 | -0.0064 | -0.0051 |
|  | (27.28)** | (23.98)** | (18.50)** | (16.54)** | (4.22)** | (2.79)** |
| DMA's | 196 | 196 | 196 | 196 | 196 | 196 |

Notes: t-statistics in parentheses (*p<.05, **p<.01). Readership data from ABC. Excludes the following 11 markets with incomplete data: Birmingham-Anniston, AL; Dallas-Fort Worth, TX; Jonesboro, AR; Joplin-Pittsburg, MO-KS; Knoxville, TN; Los Angeles, CA; Nashville, TN; New York, NY; Oklahoma City, OK; Tulsa, OK; and Juneau, AK.

## 6. Conclusion

The analyses above demonstrate that increases in ownership concentration lead firms to differentiate products to a greater extent and cover a larger number of reporting beats. Moreover, this additional coverage may extend markets to new readers. Since new readers that enter the market are better off, existing readers benefit from additional choice, and prices do not rise with concentration, consumers do not appear to be harmed by consolidation.

With respect to current policy, results presented here suggest that government intervention to increase the number of media products and media owners within markets may be unnecessary. To the extent that policy is concerned with aspects of diversity other than those associated with content variety, these results identify a benefit of concentration against which other potential costs should be weighed.[37] However evidence in this paper challenges the notion that preserving multiple viewpoints necessarily makes consumers better off. If redundant coverage is valuable, then loss of owners in a market should reduce total readership. Since there is no evidence for readership decline, individuals do not appear to be made worse off by consolidation. Arguments that ownership diversity generates external political and social benefits depend ultimately on consumption as well, hence the presence of externalities is not sufficient to justify intervention.[38]

---

[37] Advocates of strong media regulation often distinguish variety in content or programming from viewpoint diversity. See, for example, the response to FCC's Notice of Inquiry on the Commission's Newspaper/Radio Cross-Ownership Waiver Policy by a coalition of minority interest groups, *Comments Of Black Citizens For A Fair Media et al.*, MM Docket No. 96-197 dated February 7, 1997.

[38] Because it is not possible generally to measure infra-marginal utility, the potential for increases in ownership concentration to reduce total consumer surplus while increasing readership cannot be ruled out. This might occur, for example, if new coverage replaces existing material from which certain types of consumers derive large benefits. One way to check whether a particular group is harmed by concentration while others benefit is to regress changes in readership across race on changes in ownership concentration. Ownership concentration does not appear to differentially affect readership among blacks relative to whites.

However despite evidence that ownership concentration has not harmed consumers, it is not possible to conclude from this research alone that current policies are misguided. First, this paper only considers the effect of concentration on consumers. Nothing can be said about aggregate welfare without taking into account how ownership concentration affects advertising prices, hence policies limiting consolidation may still be warranted when advertiser welfare is taken into account. It also might be the case that increases in ownership concentration and coverage in newspaper markets in the 1990's are related to heightened competition with radio and cable television over this period. Although results in this paper are consistent with findings in the literature that ownership concentration in radio produces greater programming variety, little work has been done to directly examine competition across media and it remains an important area for further research.

In sum, regulation of media markets in the U.S. and antitrust policy in particular presume that more owners and more products lead to greater content variety. However the effect of concentration on variety in differentiated product markets is an empirical question that depends on fixed costs and the value of consumers to advertisers. The analyses presented here suggest that concentration in newspaper markets does not, in fact, harm consumers. Policies that prevent consolidation in media markets may thus be unwarranted.

# APPENDIX

**Table A1: Number of Markets Covered by Beat, 1993 & 1999 (All Beats)**

| Beat | 93 | 99 | Beat | 93 | 99 |
|---|---|---|---|---|---|
| Administration | 0 | 19 | Conservation & Environment-Wildlife | 0 | 1 |
| Advertising | 204 | 201 | Consumer Interests-General | 99 | 75 |
| Advertising & Public Relations-General | 0 | 29 | Creative / Graphics | 9 | 33 |
| Aeronautics & Astronautics-Aviation | 3 | 8 | Criminology & Law Enforcement-General | 0 | 24 |
| Aeronautics & Astronautics-General | 0 | 9 | Disability & Physically Challenged-General | 0 | 1 |
| Agriculture-General | 158 | 140 | Drugs & Pharmaceuticals-General | 0 | 1 |
| Agriculture-Tobacco | 0 | 1 | Education-College & Post-Graduate | 0 | 10 |
| Apparel & Accessories-General | 197 | 180 | Education-General | 179 | 168 |
| Architecture-General | 0 | 11 | Electricity & Electronics-Consumer | 0 | 6 |
| Art & Sculpture-General | 0 | 21 | Electricity & Electronics-General | 0 | 4 |
| Automotive Industry-General | 146 | 126 | Energy-Trade | 0 | 17 |
| Automotive Industry-Motorcycle & Truck | 0 | 1 | Entertainment-General | 196 | 196 |
| Banking & Finance-Consumer Finance | 0 | 27 | Entertainment-Movies, Video, TV & Radio | 148 | 164 |
| Banking & Finance-General | 155 | 116 | Entertainment-Theater & Performing Arts | 139 | 130 |
| Beverage Industry-Wine & Wineries | 0 | 5 | Entertainment-TV/Radio Listing Guides | 0 | 18 |
| Building, Construction & Demolition-Trade | 1 | 2 | Ethnic & Multicultural-General | 0 | 11 |
| Business & Economy - Employment | 0 | 14 | Ethnic & Multicultural-Hispanic | 0 | 3 |
| Business & Economy -Retail | 2 | 23 | Finance | 0 | 7 |
| Business & Economy -Economic Conditions | 0 | 18 | Fitness, Health & Hygiene-General | 121 | 130 |
| Business & Economy -International Trade | 0 | 11 | Food & Grocery Trade | 0 | 2 |
| Business & Economy -Investments | 0 | 21 | Forestry-Lumber & Wood | 0 | 1 |
| Business & Economy -Management Pubs. | 0 | 1 | Gardening & Horticulture-Trade | 108 | 101 |
| Business & Economy -National | 205 | 193 | General Interest-Advice | 0 | 1 |
| Business & Economy -Non-Profits | 0 | 1 | General Interest-Bridal & Marriage | 0 | 1 |
| Business & Economy -Regional | 0 | 11 | General Interest-Children | 74 | 65 |
| Business & Economy -Small Business | 0 | 25 | General Interest-Dining & Cooking | 199 | 195 |
| Business & Economy -Taxation | 0 | 2 | General Interest-Gossip | 0 | 1 |
| Child Care & Child Development | 115 | 96 | General Interest-Home | 164 | 156 |
| Circulation | 203 | 199 | General Interest-National | 204 | 205 |
| Civil Engineering-General | 0 | 1 | General Interest-Regional & Local | 0 | 28 |
| Classified | 184 | 172 | General Interest-Singles & Dating | 0 | 1 |
| Collecting-Antiques | 0 | 2 | General Interest-Women | 165 | 140 |
| Collecting-Stamps & Currency | 0 | 1 | General Interest-Young Adult | 1 | 3 |
| Communications | 0 | 2 | Government & Politics-General | 158 | 148 |
| Communications-Broadcasting-Cable | 0 | 4 | Government & Politics-International | 0 | 3 |
| Communications-Broadcasting-Radio-TV | 3 | 39 | Government & Politics-Local & State | 6 | 30 |
| Communications-General | 0 | 2 | History & State History-General | 0 | 3 |
| Communications-Multimedia | 0 | 2 | Home Improvement-Home Furnishings | 0 | 1 |
| Communications-Telecommunications | 0 | 10 | Hospitals & Healthcare Management | 0 | 14 |
| Community Relations | 2 | 8 | Insurance-General | 0 | 8 |
| Computers & Computerization-General | 98 | 99 | Labor & Labor Unions-General | 72 | 62 |
| Computers & Computerization-Info. Mgt. | 0 | 1 | Law-Courts Reporter | 0 | 22 |
| Conservation & Environment-General | 119 | 125 | Law-General | 88 | 88 |

| Beat | 93 | 99 | Beat | 93 | 99 |
|---|---|---|---|---|---|
| Letters to the Editor | 0 | 12 | Religions & Theology-General | 145 | 141 |
| Librarian / Research | 1 | 13 | Restaurant & Hotel Management-General | 0 | 2 |
| Literature-Book Reviews | 177 | 167 | Sales | 9 | 22 |
| Literature-General | 0 | 1 | Sciences-Archaeology | 0 | 1 |
| Literature-Humor & Satire | 0 | 7 | Sciences-Astrology & Parapsychology | 0 | 1 |
| Management | 206 | 207 | Sciences-Astronomy | 0 | 2 |
| Manufacturing, Machinery & Equipment | 0 | 16 | Sciences-General | 127 | 105 |
| Mass Transportation & Shipping-General | 58 | 46 | Sciences-Meteorology | 0 | 1 |
| Mass Transportation & Shipping-Railroads | 0 | 2 | Senior Citizens-General | 0 | 6 |
| Letters to the Editor | 0 | 12 | Social Sciences & Sociology-General | 0 | 8 |
| Media Relations | 0 | 3 | Social Service & Welfare-Counseling | 0 | 1 |
| Medicine-AIDS/HIV | 0 | 1 | Social Service & Welfare-General | 0 | 5 |
| Medicine-General | 146 | 140 | Social Service & Welfare-Subst. Abuse | 0 | 1 |
| Medicine-Psychology | 0 | 1 | Sports-Auto Racing | 0 | 7 |
| Meetings & Conventions-General | 0 | 1 | Sports-Baseball | 0 | 5 |
| Military & Defense Industry-General | 0 | 9 | Sports-Bicycling | 0 | 1 |
| Music-Classical, Choral & B& Music | 9 | 30 | Sports-Boats & Boating | 0 | 2 |
| Music-Country, Folk & Bluegrass Music | 0 | 1 | Sports-Bowling | 0 | 2 |
| Music-General | 129 | 127 | Sports-General | 205 | 226 |
| Music-Jazz & Blues | 1 | 4 | Sports-Golf | 0 | 27 |
| Music-R & B, Urban, World & Latin Music | 0 | 1 | Sports-Horses & Horsemanship | 0 | 10 |
| Music-Rock Music | 10 | 33 | Sports-Ice Skating & Hockey | 0 | 5 |
| New Media | 0 | 31 | Sports-Skiing & Snow Sports | 0 | 1 |
| News-Foreign | 62 | 48 | Sports-Soccer & Rugby | 0 | 3 |
| News-General | 182 | 207 | Sports-Sports Industry & Business | 0 | 6 |
| News-Local | 2 | 38 | Sports-Sports Teams | 0 | 17 |
| News-National | 74 | 86 | Sports-Tennis & Racquet Sports | 0 | 8 |
| News-Regional | 192 | 187 | Sports-Track & Field, Gymnastics | 0 | 2 |
| Operations | 1 | 15 | Technology-General | 113 | 84 |
| Opinion & Commentary | 199 | 196 | Travel Industry-Consumer | 181 | 175 |
| Pets-Dogs | 0 | 1 | Women's Specialty & Feminist-General | 0 | 2 |
| Photo | 174 | 169 | | | |
| Polls | 0 | 3 | | | |
| Printing & Graphic Arts-General | 0 | 1 | | | |
| Product Development | 0 | 2 | | | |
| Production | 37 | 23 | | | |
| Promotion / Marketing | 33 | 54 | | | |
| Publishing-Newspaper Publishing & Journalism | 0 | 2 | | | |
| Real Estate-General | 154 | 142 | | | |
| Recreation, Leisure & Amusement-Consumer | 0 | 2 | | | |
| Recreation, Leisure & Amusement-Gambling | 0 | 3 | | | |
| Recreation, Leisure & Amusement-Outdoor | 181 | 171 | | | |

**Table A2: Retail Summary Statistics, 1997 Economic Census**

| Retail Category | NAICS Code | Mean Stores (DMA) | Mean Sales (000) |
|---|---|---|---|
| Furniture | 44211 | 20.50 | 33,544 |
| Floor covering | 44221 | 24.01 | 31,511 |
| Other home furnishings | 44229 | 26.42 | 27,465 |
| Appliance, TV, Electronics | 44311 | 22.60 | 51,414 |
| Computer and software | 44312 | 12.44 | 56,713 |
| Camera and photographic supplies | 44313 | 6.38 | 11,072 |
| Home centers | 44411 | 5.02 | 153,889 |
| Paint and wallpaper | 44412 | 14.31 | 24,152 |
| Hardware | 44413 | 19.65 | 27,373 |
| Other building material dealers | 44419 | 27.20 | 89,233 |
| Outdoor power equipment | 44421 | 7.37 | 10,012 |
| Nursery and garden centers | 44422 | 7.13 | 17,309 |
| Supermarkets | 44511 | 69.92 | 399,043 |
| Convenience | 44512 | 52.28 | 36,240 |
| Pharmacies and drug | 44611 | 66.43 | 180,486 |
| Cosmetics, beauty, perfume | 44612 | 12.26 | 9,003 |
| Optical goods | 44613 | 29.95 | 14,938 |
| Other health and personal care | 44619 | 22.05 | 16,910 |
| Men's clothing | 44811 | 25.93 | 27,482 |
| Women's clothing | 44812 | 88.41 | 74,818 |
| Children's and infants' clothing | 44813 | 10.38 | 18,011 |
| Family clothing | 44814 | 30.37 | 86,698 |
| Clothing accessories | 44815 | 12.48 | 8,498 |
| Other clothing | 44819 | 18.92 | 15,439 |
| Shoe | 44821 | 58.15 | 44,271 |
| Jewelry | 44831 | 50.08 | 48,213 |
| Luggage and leather goods | 44832 | 4.88 | 8,823 |
| Sporting goods | 45111 | 38.96 | 41,201 |
| Hobby, toy, and game | 45112 | 18.78 | 34,701 |
| Sewing, needlework, and piece goods | 45113 | 10.23 | 10,080 |
| Musical instrument and supplies | 45114 | 7.95 | 13,935 |
| Book and news dealers | 45121 | 18.73 | 29,890 |
| Prerecorded tape, CD and records | 45122 | 14.99 | 19,499 |
| Department Stores | 45211 | 17.84 | 459,330 |
| Warehouse clubs | 45291 | 3.22 | 347,130 |
| All other general merchandise | 45299 | 11.50 | 8,156 |
| Florists | 45311 | 40.13 | 12,400 |
| Office supplies and stationery | 45321 | 6.65 | 37,262 |
| Gift, novelty, and souvenir | 45322 | 56.61 | 28,100 |
| Used merchandise | 45331 | 22.30 | 10,020 |
| Pet and pet supplies | 45391 | 13.25 | 13,223 |
| Art dealers | 45392 | 10.49 | 10,885 |
| Manufactured (mobile) home dealers | 45393 | 5.40 | 18,909 |
| All other miscellaneous store retailers | 45399 | 15.58 | 22,634 |